\documentclass[iop]{emulateapj}
\usepackage{graphicx}
\usepackage{adjustbox}

\newcommand{\Vt}{$\mathrm{V_{t}}$ }

\newcommand{\Teff}{${T_{\rm eff}}$ }

\newcommand{\logg}{$\log \mathrm{g}$ }

\received{\today}

\shorttitle{Metal-Poor Type II Cepheids }
\shortauthors{Kovtyukh et al.}

\begin{document}

\title{Metal-Poor Type II Cepheids with Periods Less Than Three Days}

\author{V. Kovtyukh}
\affiliation{Astronomical Observatory, Odessa National University, Shevchenko Park, 650014, Odessa, Ukraine}
\affiliation{Isaac Newton Institute of Chile, Odessa Branch, Shevchenko Park, 650014, Odessa, Ukraine}
  
\author{G. Wallerstein}
\affiliation{Astronomy Department, University of Washington, Seattle, WA 98105, USA}

\author{I. Yegorova}
\affiliation{Universidad Andres Bello, Fernandez Concha 700, Santiago, Chile}

\author{S. Andrievsky}
\affiliation{Astronomical Observatory, Odessa National University, Shevchenko Park, 650014, Odessa, Ukraine}
\affiliation{Isaac Newton Institute of Chile, Odessa Branch, Shevchenko Park, 650014, Odessa, Ukraine}
\affiliation{GEPI, Observatoire de Paris-Meudon, CNRS, Universite Paris Diderot, 92125, Meudon Cedex, France}

\author{S. Korotin}
\affiliation{Astronomical Observatory, Odessa National University, Shevchenko Park, 650014, Odessa, Ukraine}
\affiliation{Crimean Astrophysical Observatory, Nauchny 298409, Crimea, Ukraine}

\author{I. Saviane}
\affiliation{European Southern Observatory, Alonso de Cordova 3107, Santiago, Chile}

\author{S. Belik}
\affiliation{Astronomical Observatory, Odessa National University, Shevchenko Park, 650014, Odessa, Ukraine}
\affiliation{Isaac Newton Institute of Chile, Odessa branch, Shevchenko Park, 650014, Odessa, Ukraine}

\author{C. E. Davis}
\affiliation{Astronomy Department, University of Washington, Seattle, WA 98105, USA}

\author{E. M. Farrell}
\affiliation{Astronomy Department, University of Washington, Seattle, WA 98105, USA}

\begin{abstract}

We have analysed 10 high resolution spectra of Type II Cepheids with periods less than 3 days. We find that they clearly separate into two groups: those with near or slightly below solar metallicities, and those with [Fe/H] between --1.5 and --2.0. While the former are usually called BL~Her stars, we suggest that the latter be called UY~Eri stars. The UY~Eri subclass appears to be similar to the short period variables in globular clusters of the Galactic Halo. Globular clusters with [Fe/H] $\textgreater$ --1.0 almost never have Type II Cepheids.

\end{abstract}


\section{Introduction}
\label{sec:intro}

The origin of the galactic halo has been an interesting question ever since halo stars and clusters became the defining objects of population II (\citealt{Baade1944}, \citealt{OConnell1958}). Halo subdwarfs were soon recognized to be metal-poor (\citealt{Chamberlain1951}), as were red giant stars branch (RGB) stars in glubular clusters of the halo (\citealt{Helfer1959}). Halo cepheids are now called Type II Cepheid (T2C) stars, and were first recognized by \citet{Joy1937}, and found to be metal-poor by \citet{Abt1954} and \citet{Rodgers1963}. Short period cepheids were found in metal-poor globular clusters by \citet{Arp1955}, and catalogued by \citet{Clement2001}.\footnote{The most recent update is to be found at http://www.astro.utoronto.ca/~cclement/cat/listngc.html}

To the surprise of some, \citet{Woolley1966} found that the halo cepheids showed thick disk kinematics and, hence, were likely to be only moderately metal-poor. The short period stars received the name BL~Her, a star which has a period of 1.3 days, and is known to show a metallicity of --0.16 (\citealt{Maas2007}).

In an effort to further understand the T2C stars of the galactic halo, \citet{Maas2007} derived metallicities of 19 stars, 7 of which have periods of 3 days or less. Except of one star, UY~Eri, they showed very modest depletions of heavy elements and, hence, fit the classification of BL~Her stars. To further expand the database for T2C stars in the halo, we have obtained high resolution spectra of field T2C stars, 10 of which have periods below 3 days. The stars were selected to be bright enough for the available telescopes, and to be conveniently placed when observing time was assigned.

In this paper, we report on the chemical composition of the 10 T2C field stars with periods from 0.9 to 3 days. The upper limit marks the edge of an almost empty gap from 4 to 9 days in the distribution shown in Figure~1 of \citet{Soszynski2008} and Figure~6 of \citet{Schmidt2011}. The separation of both cepheid strip candidates and Type II Cepheids into metal-normal and metal-poor stars can be seen in Figure~8 of \citet{Schmidt2011}.

\begin{table*}[t]
    \scriptsize
    \caption{Details of Observations and Stellar Parameters}
    \label{tab:atmparam}
    \begin{adjustbox}{width=\textwidth,center=\textwidth}
    {\begin{tabular}{rcccrccccrl}
    \hline \hline
    Star & P(days) & Telescope & JD (2400000+) & Exp (s) & Phase & \Teff{} (K) & \logg{} & \Vt{} (km s$^{-1}$) & [Fe/H] & Remarks \\
    \hline
    UY~CrB & 0.929 & APO & 53162.6736 & 3600 & 0.491 & 6150 & 1.8 & 2.0 & --0.43 & \\
     & & APO & 53456.9981 & 1800 & 0.245 & 6700 & 2.0 & 2.4 & --0.32& \\
     & & APO & 53626.7528 & 1800 & 0.839 & 6300 & 2.5 & 2.0 & --0.47& \\
     NSV 10788 & 1.081 & VLT & 56124.7937 & 1200 & 0.451 & 6250 & 2.3 & 2.8 & --2.41 & \\
     V716~Oph & 1.116 & APO & 52336.9382 & 1200 & 0.154 & 7000 & 1.8 & 2.7 & --1.64 & \\
     & & APO & 52417.7674 & 1800&0.595 & 6100 & 2.2 & 2.6 & --1.67 & \\
     & & APO & 52447.6805 & 1800&0.380 & 6600 & 2.6 & 2.6 & --1.56 & \\
     & & APO & 52448.6667 & 1800&0.284 & 6700 & 2.2 & 2.6 & --1.72 & \\
     & & APO & 52449.7423 & 1800&0.248 & 6750 & 2.0 & 2.3 & --1.62 & \\
     BF~Ser & 1.165 & APO & 52336.8431 & 2400 & 0.652 & 5800 & 1.0 & 2.3 & --2.15 & \\
     & & APO & 52417.7090 & 1800 & 0.038 & 7300 & 2.2 & 3.0 & --2.04 & \\
     & & APO & 52804.7590 & 900 & 0.145 & 7000 & 2.0 & 3.0 & --2.08 & H$\alpha$ double or emission \\
     & & APO & 53457.8495 & 1800 & 0.526 & 6300 & 2.1 & 2.2 & --2.04 & H$\alpha$ double or emission \\
     BL~Her & 1.307 & APO & 53163.8646 & 900 & 0.104 & 7000 & 2.2 & 2.2 & --0.12 &    \\
     & & OHP & 49572.3784 & 900 & 0.147 & 6650 & 2.5 & 2.2 & --0.20 &   \\
     XX~Vir & 1.348 & APO &52417.6479 & 1800 & 0.070 & 7500 & 2.2  & 2.3 & --1.62& \\
     & & APO &52449.7188 & 1800 & 0.858 & 6100 & 2.5  & 2.8 & --1.51& \\
     & & APO &53541.6750 & 1800 & 0.791 & ... & ... & ... & ... &   \\
     V1287~Sco & 1.956 & VLT &56126.4862 & 1500 & 0.487 & 5950 & 2.2 & 3.5 & --1.94 & H$\alpha$ emission \\
     V553~Cen & 2.061 & ESO & 56748.6532 & 90 & 0.818 & 6060 & 2.2 & 2.7 & 0.01 & see \citet{Wallerstein1996} \\
     UY~Eri & 2.213 & APO & 52984.7985 & 1800 & 0.889 & 6400 & 2.0 & 2.0 & --1.83 & H$\alpha$ emiss or double \\
     & & APO & 53687.7856 & 1800 & 0.511 & 6200 & 1.8 & 2.6 & --1.66 & \\
     & & APO & 54044.7828 & 1800 & 0.809 & 6000 & 1.9 & 2.6 & --1.70 & H$\alpha$ emission \\
     AU~Peg & 2.402 & APO & 53626.7292 & 900 & 0.848 & 6008 & 2.0 & 2.8 & 0.33 & orbit phase =  0.538 \\
     & & APO & 53687.6390 & 900 & 0.109 & 5544 & 1.5 & 2.3 & 0.21 & orbit phase = 0.681 \\
    \hline
    \end{tabular}}
    \end{adjustbox}
    \begin{itemize}
    \item[] {\it Remarks: \\}
    
(a) -- {\it NSV 10788} is a star with P = 1.08 days and a metallicity of [Fe/H] = --2.4. The lightcurve shows an amplitude of 0.5 magnitudes. Thus, classification of this star as a cepheid is uncertain.

(b) -- For {\it BF~Ser}, all four spectra were taken at important phases. This star is very likely a member of the UY~Eri group. It is a metal poor-star with a pulsation period of 1.2 days.

(c) -- {\it XX~Vir} is a high latitude star of the UY~Eri group with a short pulsation period of 1.3 days. For one spectrum, with emission in iron lines, the chemical composition was not determined.

(d) --  {\it V1287~Sco} is a variable star of UY~Eri type. There is not much information in the literature about this star. H$\alpha$ emission is seen in the spectrum. The lines are slightly asymmetric.

(e) -- {\it V553~Cen} is a star recognized by \citet{Evans1983} as a rare C-rich Cephied. Its composition was investigated by \citet{Wallerstein1998}.

(f) -- {\it AU~Peg} is a spectroscopic binary \citep{Harris1984} with an orbital period of 53.3 days. The chemical composition of this star was studied by \citet{Harris1984} and \citet{Maas2007}. It appears to be slightly metal rich, which may be due in part to mass transfer from its unobserved companion.

    \end{itemize}
\end{table*}

\section{Observations and Data Reduction}
\label{sec:obs}

Seven objects were observed with the 3.5-m telescope at the Apache Point Observatory with the ARC Echelle Spectrograph (ARCES). By using a prism as a cross-disperser, the APO echelle captures the entire spectrum from 3500 \AA{} to 10\,400 \AA{} with a resolving power of 31\,500. However, the red-sensitive 2048x2048 chip has decreasing sensitivity for cool stars at shorter wavelengths and beyond 9000 \AA{}. The observations were obtained as a part of a program to derive the chemical composition of certain Type II Cepheids and RR Lyrae stars. The exposure times were approximately 10--30 minutes. The estimated S/N ratio per pixel at the continuum level, depending upon the wavelength interval, is approximately 80--150. The uncertainty in the determination of velocities is a few tenths of a km s$^{-1}$.

Two objects were observed with the cross-dispersed echelle spectrograph, UVES, at the Very Large Telescope.\footnote{Based on observations collected at the European Organization for Astronomical Research in the Southern Hemisphere under ESO programme 089.D-0489(A).} The red arm was used, which covers the wavelength region between 4200 \AA{} and 11\,000 \AA{}. Standard instrumental settings were used to achieve wavelength coverage from 4790--5760 \AA{} and 5830--6810 \AA{} with a resolution of 0.16 \AA{}. The observations were done in service mode. For some objects, several spectra were obtained. The exposure times were approximately 20--30 minutes. The primary data reduction, such as bias subtraction, flat-field correction, wavelength calibration, sky subtraction and spectra extraction was performed with the UVES pipeline (\citealt{Ballester2000}).

One object, the carbon cepheid, V553~Cen, was taken from the ESO archive. The star was observed with the echelle spectrograph HARPS at the ESO La Silla 3.6-m telescope. The spectral range is 4000--6800 \AA{} with a resolving power of 100\,000. For BL~Her, an additional spectrum was taken from the archive of the Elodie spectrograph at the Observatoire de Haute-Provence 1.93-m telescope (R=40\,000, $\lambda$ 4000-6800 \AA{}). The observed stars and their atmospheric parameters are given in Table 1.

\section{Spectroscopic Analysis}
\label{sec:spec}

To determine the effective temperature, T$_{eff}$, we used the line depth ratio method of \citet{Kovtyukh2007}. The uncertainties in this method range from 30--100 K depending on the S/N. To set the $\log{g}$ value, we required that the iron abundance, as determined from lines of FeI and FeII, be equal. 

Elemental abundances were determined using LTE and NLTE approximations combined with atmospheric models by \citet{Castelli2004}, computed for the parameters of each star. The solar abundances were computed for lines from the solar spectrum \citep{Kurucz1984} with $\log{gf}$ from the Vienna Atomic Line Database (VALD) \citep{Kupka1999}, and the solar model by \citet{Castelli2004}. They are listed in \citet{Lemasle2015}. 

\begin{table}
    \centering
    \caption{Table of $\log (gf)$ and $\chi_{\rm{exc}}$ for the Na and O lines used.}
    \label{tab:loggf}
    \begin{tabular}{crr}
    \hline \hline Lambda & $\chi_{\rm{exc}}$ & $\log (gf)$\\ \hline
    & Sodium & \\ \hline
    5682.63 & 2.09 & --0.71 \\
    5688.19 & 2.10 & --0.41 \\
    6154.23 & 2.09 & --1.56 \\
    6160.75 & 2.10 & --1.26 \\
    \hline
    & Oxygen & \\ \hline
    6300.30 & 0.00 & --9.717 \\
    7771.94 & 9.11 & 0.369 \\
    7774.17 & 9.11 & 0.223 \\
    7775.39 & 9.11 & 0.001 \\
    8446.25 & 9.52 & --0.462 \\
    8446.36 & 9.52 & 0.236 \\
    8446.76 & 9.52 & 0.014 \\ 
    \hline
    \end{tabular}
\end{table}

\subsection{Oxygen}

The NLTE model of the oxygen atom was first described by \citet{Mishenina2000}, and then updated by \citet{Korotin2014}. The model consists of 51 OI levels of singlet, triplet, and quintet systems, and the ground level of the OII ion. Fine structure splitting was taken into account only for the ground level and the 3p5P level (the upper level of the 7772,4,5 triplet lines). A total of 248 bound-bound transitions were included. Oxygen line parameters are listed in Table \ref{tab:loggf}. The high excitation OI triplet suffers from departure from LTE (\citealt{Parsons1964}, \citealt{Altrock1968}, \citealt{Amarsi2016}). Its strength depends sensitively on the star's surface gravity. In stars of high luminosity, the triplet is greatly enhanced since radiative effects dominate recombination and ionization, as compared to collisional excitation, which is controlled by the local temperature.

\subsection{Sodium}

Sodium line parameters are given in Table \ref{tab:loggf}. We derived the sodium abundances by line profile fitting. The NLTE atomic model of sodium was presented by \citet{Korotin1999} and then updated by \citet{Dobrovolskas2014}. The updated sodium model currently consists of twenty energy levels of NaI and the ground level of NaII. In total, 46 radiative transitions were taken into account for the calculation of the population of all levels. For four stars, the Na D lines were saturated, so we used the pair of lines at 5682 \AA{} and 5688 \AA{}. For stars with nearly solar metallicity, we used the weaker pair at 6154 \AA{} and 6160 \AA{}. We chose not to use the 8183 \AA{} and 8194 \AA{} lines due to blending with absorption by atmospheric lines, which depends on the stellar radial velocity and the humidity above the observatory.

The results of the abundance analysis on our program stars is given in Table \ref{tab:MetalCMn}.

\section{Discussion}
\label{sec:disc}

The T2C stars have been divided into W Vir stars with periods from 9 to 30 Days (though stars with periods greater than 20 days often show mild RV Tau properties); and the BL Her stars with periods less than 3 days.\footnote{The gap from 3 to 9 days is violated by variable 3 in the globular cluster M10.} We advocate for the  separation of the 1--3 day stars into the BL Her group with near-solar metallicity, and the distinctly metal-poor group that we will call the UY Eri class. The latter are to be found also in metal-poor globular clusters \citep{Clement2001}. The relatively metal-rich globulars with [Fe/H] $\textgreater$ --1.0 do not have cepheids with periods less than 3 days. Of 10 stars in Table 1, 6 fall into the metal poor UY Eri group, all of which show [Fe/H] $\textless$ --1.5.

Our database may be enhanced by including the abundances in \citet{Maas2007}. Just by chance, all of their short period stars, except for UY~Eri, fall into the more metal-rich BL~Her class. Hence, we have combined the two sources to develop mean abundance ratios for the BL~Her subgroup, and used only our data for the metal-poor UY~Eri group. Our abundance data are summarized in Table~\ref{tab:OurStats}.

\begin{figure}
\epsscale{1.2}
\plotone{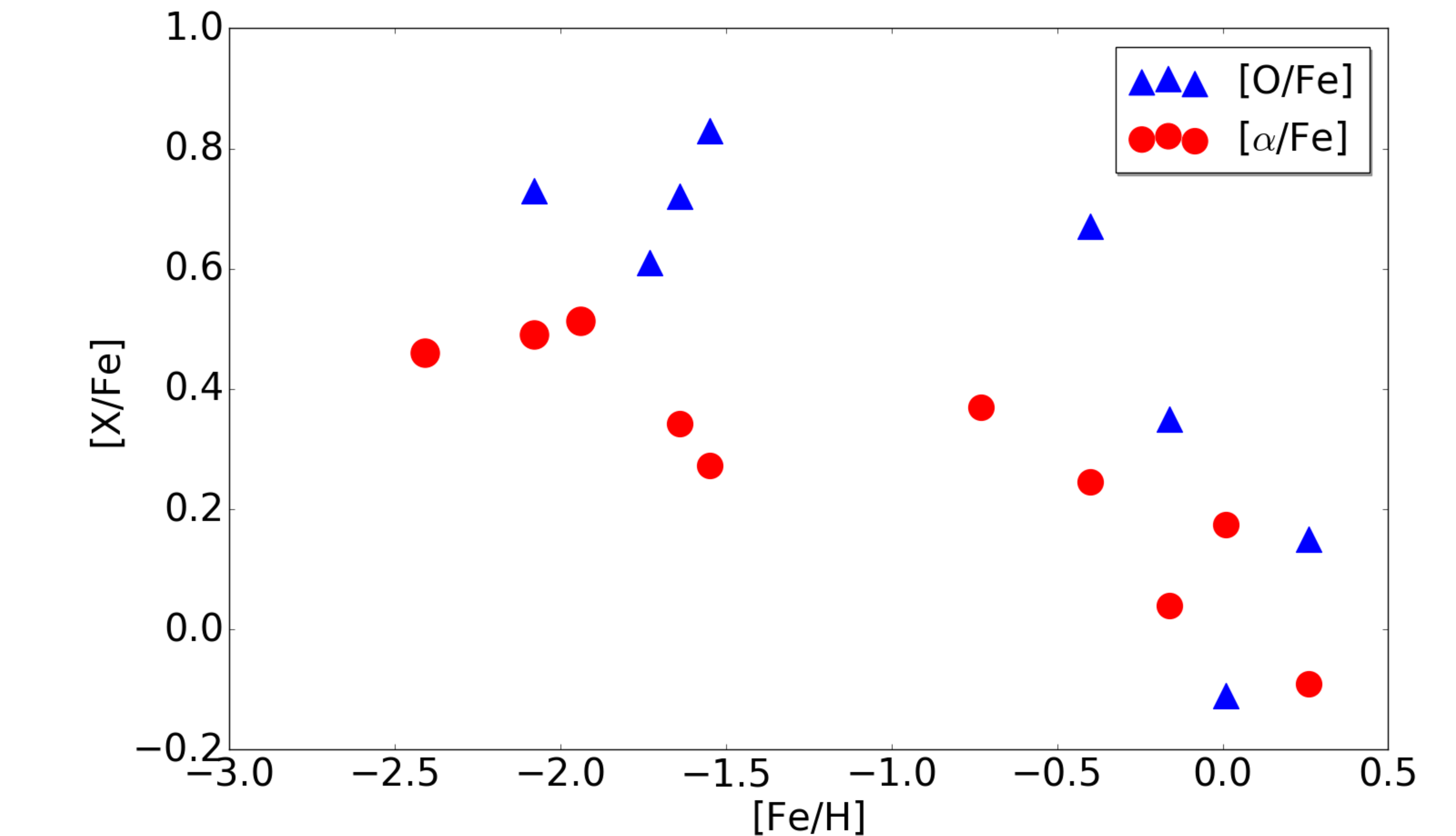}
\caption{~The dependence of [O/Fe] and [$\alpha$/Fe] on [Fe/H] for the observed stars. \label{XFe}}
\end{figure}

For our small sample, formal probable errors are less indicative of the accuracy of a quantity than one might hope. Hence, we show the dependence of [O/Fe] and the dependence of [$\alpha$/Fe] on [Fe/H] in Figure~\ref{XFe}, using the elements Mg, Si, Ca, and Ti. The dependence of [O/Fe] on [Fe/H] shows a dependence that is similar to what is seen in both giants and dwarfs, with a rise of [O/Fe] from near 0.0 at [Fe/H] = 0.0 to +0.7 between [Fe/H] = --1.5 to --2.5. For [$\alpha$/Fe], the pattern is similar to many other investigations of the $\alpha$-elements, rising from near 0.0 to +0.5 for the metal-poor UY~Eri Group.

The metal-rich BL Her group show abundances of 22 elements that differ little in their ratios relative to iron, with the possible exception of sodium. For the UY~Eri group, the abundances of oxygen, and the $\alpha$-elements from Mg to Ti  show excesses that are comparable to the excesses in most metal-poor red giants and main sequence stars.

We await the discovery and observation of additional T2C stars by Gaia and eventually the Large Synoptic Survey Telescope (LSST). The Gaia spectra in the 8400--8800 \AA{} region may be used to derive metallicities from the strenth of the Ca II IR triplet as has been calibrated for RR Lyrae stars by \citet{Wallerstein2012}.

\subsection{Carbon, Nitrogen, and Oxygen}

There are a few CI lines of high excitation in these stars. Despite our NLTE analyses, the scatter of [C/Fe] is too great to consider the mean values of [C/Fe] to be definitive. A significant excess in [N/Fe] for the BL~Her stars is approximately what is expected from the enhancement of nitrogen by the CNO cycle and internal mixing. For the metal-poor UY~Eri stars, the nitrogen lines are too weak for analysis. For oxygen, the analysis depends upon the OI triplet at 7772,4,5 \AA{} and the 8446.63 \AA{} lines. As we have shown in Figure~\ref{XFe}, for the metal-poor UY~Eri stars, the mean excess in [O/Fe] is $0.72 \pm 0.09$, which is typical for metal-poor stars.

\subsection{Light Elements}

For sodium, we see small excesses that are approximately equal to their uncertanties for both groups of stars. In most, if not all, globular clusters, some red giants show a significant excess of sodium in second generation stars \citep{Carretta2009}. Our abundances are based on the lines arising from the 2.1 eV excited level. We did not use the Na D lines because they are usually too strong and are sometimes blended with circumstellar or interstellar lines. For the alpha elements, Mg, Si, Ca, and Ti, the BL~Her stars do not show a difference in their abundances to that of iron, as is usually seen in stars with near-solar metallicity. S is omitted from the $\alpha$-elements because its lines are too weak for analysis, except in UY~Eri. For the UY~Eri stars, we find a mean value of [$\alpha$/Fe] = $0.35\pm0.19$ which is typical of metal-poor stars.

\begin{table*}
    \tiny
    \caption{Relative to Fe Abundances in Type II Cephieds (C--Ti)}
    \label{tab:MetalCMn}
    \begin{adjustbox}{width=\textwidth,center=\textwidth}
    {\begin{tabular}{rrrrrrrrrrrrrrr}
    \hline \hline
    & P, days & [Fe/H] & C & O & Na & Mg & Al & Si & S & Ca & Sc & Ti \\
    \hline
    UY~CrB & 0.929 & --0.40 & 0.65 & 0.67 & 0.52 & 0.31 & 0.35 & 0.24 & 0.25 & 0.20 & 0.01 & 0.23 \\
    NSV 10788 & 1.081 & --2.41 & ...  & ... & ... & 0.52 & ... & ... & ... & 0.49 & 0.27 & 0.37 \\
    V716~Oph & 1.116 & --1.64 & 0.43 & 0.72 & 0.36 & 0.21 & ... & 0.14 & ... & 0.53 & 0.23 & 0.49 \\
    BF~Ser & 1.165 & --2.08 & 0.13 & 0.73 & ... & 0.31& ... & ... & ... & 0.75 & 0.20 & 0.41 \\
    BL~Her & 1.307 & --0.16 & 0.36 & 0.35 & 0.63 & 0.01 & 0.00 & --0.01 & 0.10 & 0.01 & 0.06 & 0.15 \\
    XX~Vir & 1.348 & --1.55 & ... & 0.83 & ... & 0.19 & ... & 0.05 & ... & 0.46 & 0.26 & 0.39 \\
    V1287~Sco & 1.956 & --1.94 & ... & ... & ... & 0.55 & ... & ... & ... & 0.52 & 0.25 & 0.47  \\
    V553~Cen & 2.061 & 0.01 & 0.78 & --0.11 & 0.43 & 0.19 & 0.12 & 0.00 & 0.10 & 0.23 & 0.28 & 0.28 \\
    UY~Eri & 2.213 & --1.73 & --0.19 & 0.61 & 0.18 & 0.37 & ... & 0.06 & 0.33 & 0.60 & 0.06 & 0.45 \\
    AU~Peg & 2.402 & 0.26 & 0.17 & 0.15 & --0.04 & --0.07 & 0.18 & --0.11 & 0.15 & --0.14 & --0.27 & --0.04 \\
    \hline
    \end{tabular}}
    \end{adjustbox}
    \\
\end{table*}

\section{Conclusions}
\label{sec:conc}

We have found that the T2C stars with periods of 1--3 days may be divided into a nearly normal metal group, usually called BL~Her stars, and a newly recognized group with [Fe/H] = --1.5 to --2.4. The BL~Her group probably belong to the thick disk, while latter are similar to short period cepheids in metal-poor globular clusters. With a few exceptions, it appears that globulars with [Fe/H] $\textgreater$ --1.0 do not have cepheid members of any metallicity. 

The relationship of globular clusters and the general halo of our galaxy is not clear. \citet{Martell2016} have shown that only a very small percentage of the general halo could have come from the evaporation of stars from globular clusters. Since all stars seem to form in groups and clusters, it is possible that the single halo stars are descendents of small, loose clusters no longer recognizable as such. In addition, the populations of variable stars in the halo and globulars are not the same. \\

We thank Giuseppe Bono, Charli Sakari, and Joanne Hughes for reading the manuscript and making some good suggestions. This research has been supported by the Kennilworth Fund of the New York Community Trust.

\begin{table}
    \centering
    \caption{Mean abundances of UY~Eri stars from this study, and BL~Her stars from this study and \citet{Maas2007} for critical elements relative to Fe.}
    \label{tab:OurStats}
    \begin{tabular}{lrrrrc}
    \hline \hline
    Element & Mean$_{BL}$ & $\sigma_{BL}$ & Mean$_{UY}$ & $\sigma_{UY}$ & Mean$_{\mid BL-UY \mid}$\\
    \hline
    O & 0.24 & 0.29 & 0.72 & 0.27 & 0.48 \\
    Na & 0.37 & 0.26 & 0.27 & 0.20 & 0.10 \\
    Mg & 0.08 & 0.16 & 0.35 & 0.20 & 0.27 \\
    Si & 0.02 & 0.13 & 0.13 & 0.14 & 0.11 \\
    S & 0.14 & 0.07 & 0.50 & 0.26 & 0.36 \\
    Ca & 0.05 & 0.16 & 0.54 & 0.26 & 0.49 \\
    Ti & 0.14 & 0.13 & 0.41 & 0.17 & 0.27 \\
    $[$Fe/H$]$ & --0.07 & 0.24 & --1.89 & 0.29 & 1.82 \\
    \hline
    \end{tabular}
\end{table}

\end{document}